# Large enhancement of the thermoelectric figure of merit in a ridged quantum well

Avto Tavkhelidze

Tbilisi State University, Chavchavadze ave. 13, Tbilisi 0179, Georgia

E-mail: avtotav@gmail.com

**Abstract.** Recently, new quantum features have been observed and studied in the area of ridged quantum wells (RQWs). Periodic ridges on the surface of the quantum well layer impose additional boundary conditions on electron wave function and reduce the quantum state density. As a result, the chemical potential of RQW increases and becomes ridge height dependent. Here, we propose a system composed of RQW and an additional layer on the top of the ridges forming periodic series of $p^+$–$n^+$ junctions (or metal–$n^+$ junctions). In such systems, charge depletion region develops inside the ridges and effective ridge height reduces, becoming a rather strong function of temperature $T$. Consequently, $T$-dependence of chemical potential magnifies and Seebeck coefficient $S$ increases. We investigate $S$ in the system of semiconductor RQW having abrupt $p^+$–$n^+$ junctions or metal–$n^+$ junctions on the top of the ridges. Analysis made on the basis of Boltzmann transport equations shows dramatic increase in $S$ for both the cases. At the same time, other transport coefficients remain unaffected by the junctions. Calculations show one order of magnitude increase in thermoelectric figure of merit $ZT$ relative to the bulk material.

## 1. Introduction

Quantum wells are considered the most reliable low-dimensional systems for thermoelectrics [1]. However, improvements in thermoelectric properties over bulk materials are insufficient for most applications. In this work, we present ridged quantum wells (RQWs) having advanced thermoelectric properties. RQW layer has periodic ridges on the surface. Its operation is based on the effect of quantum state depression (QSD). Periodic ridges impose additional boundary conditions on the electron wave function. Supplementary boundary conditions forbid some quantum states for free electron, and the quantum state density in the energy $\rho(E)$ reduces. According to Pauli's Exclusion Principle, electrons rejected from the forbidden quantum states have to occupy the states with higher $E$. Thus, chemical potential $\mu$ increases. In semiconductors, QSD reduces $\rho(E)$ in all energy bands including the conduction band (CB). Electrons rejected from the filled bands occupy the quantum states in the empty bands, and the electron concentration in the CB increases [2]. This corresponds to donor doping (we will refer to it as QSD doping). The QSD transfers electrons to higher energy levels. If initially the semiconductor is intrinsic, then the QSD doping will modify it to $n$-type. It is comparable with a conventional donor doping from the point of increase in $\mu$. However, there are no donor atoms. QSD doping does not introduce scattering centres and consequently allow high electron mobility. There are distinctions and similarities between the QSD forbidden quantum state and a hole. State is forbidden by the boundary conditions and cannot be occupied. However, it is not forbidden in an irreversible way. If the boundary conditions change (e.g., owing to charge depletion), then it can recombine with the electron (like the hole recombines with the electron). As the QSD forbidden state is confined to the boundary conditions (macroscopic geometry), it is not localized in the lattice and cannot move like a hole.

Density of states in RQW (figure 1) reduces $G$ times $\rho(E) = \rho_0(E)/G$, where $\rho_0(E)$ is the density of states in a conventional quantum well layer of thickness $L$ $(a = 0)$ and $G$ is the geometry factor. In the first approximation, for the case $L, w >> a$ and within the range $5 < G < 10$, the following simple expression can be used

$$G \approx L/a \quad . \tag{1}$$

where $a$ is the ridge height and $L$ is the RQW layer thickness (figure 1). Density of QSD

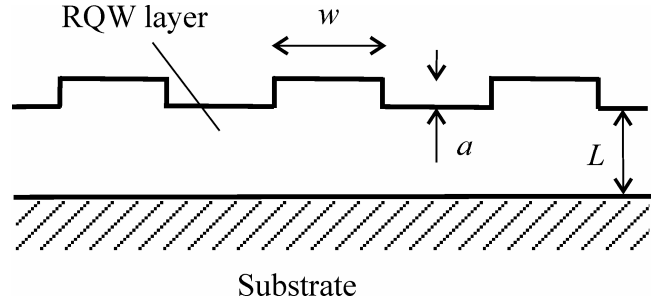

**Figure 1.** Cross-section of a ridged quantum well.

forbidden quantum states is

$$\rho_{FOR}(E) = \rho_0(E) - \rho_0(E)/G = \rho_0(E)(1 - G^{-1}) . \tag{2}$$

To determine the number of rejected electrons $n$, (2) should be integrated over electron confinement energy range

$$n = \int_{CON} dE \, \rho_{FOR}(E) = (1 - G^{-1}) \int_{CON} dE \, \rho_0(E) = (1 - G^{-1}) n_{CON} \quad . \tag{3}$$

Here, $n_{CON} = \int_{CON} dE \, \rho_0(E)$ is the number of quantum states (per unit volume) within electron confinement energy range (which depends on RQW and substrate band structures). RQW retains quantum properties at $G$ times more widths with respect to the conventional quantum well. Previously, QSD was studied experimentally [3] and theoretically [4] in ridged metal films.

Thermoelectric materials are characterized in terms of dimensionless figure of merit $ZT$ [5]. Here, $T$ is the temperature and $Z$ is given by $Z = \sigma S^2 /(\kappa_e + \kappa_l)$, where $S$ is the Seebeck coefficient, $\sigma$ is electrical conductivity, $\kappa_e$ is electron gas thermal conductivity, and $\kappa_l$ is lattice thermal conductivity. The difficulty in increasing $ZT$ is that materials having high $S$ usually have low $\sigma$. When $\sigma$ is increased, it leads to an increase in $\kappa_e$, following Wiedemann–Franz law, and $ZT$ does not improve much. Another approach is to eliminate the lattice thermal conductivity by introducing vacuum nanogap between the hot and cold electrodes [6–8] and using electron tunnelling. Cooling in such designs was observed in [9] and theoretically studied in [10, 11]. However, vacuum nanogap devices appear extremely difficult to fabricate. In this work, we present a solution that allows large enhancement of $S$ without changes in $\sigma$, $\kappa_e$, and $\kappa_l$. It is based on RQW having series of p$^+$–n$^+$ or metal–n$^+$ junctions on the top of the ridges RQW (figure 2). Depletion region width $d(T)$ depends rather strongly on temperature. The ridge effective height $a_{eff}(T) = a - d(T)$ and consequently the geometry factor of RQW becomes temperature-dependent, $G = G(T)$. All parameters of RQW including $\mu$ become stronger functions of $T$ than it will be in an RQW without the junctions. Seebeck coefficient and thermoelectric figure of merit increase.

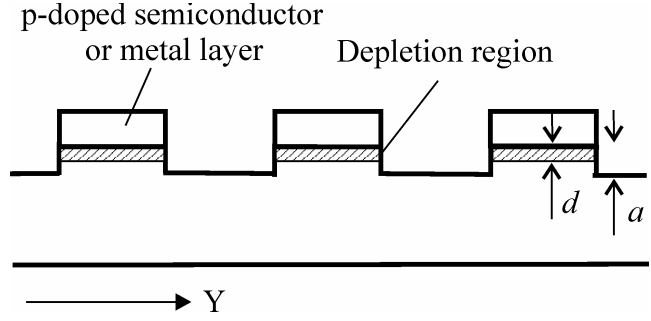

**Figure 2.** RQW with series of periodic junctions grown at the top of the ridges.

The objective of this work is to calculate $ZT$ of RQW with series of p$^+$–n$^+$ and metal–n$^+$ junctions and compare it with $Z_0T$ of reference RQW (RRQW), in which $G \neq G(T)$, and bulk material. Analysis was made using the Boltzmann transport equations. First, we calculate $\nabla\mu$ for the system of RQW and junctions and express it as

$$\nabla\mu = \nabla\mu_0 + \nabla\mu_J . \tag{4}$$

Here, $\nabla\mu_0$ is the chemical potential gradient in an RRQW and $\nabla\mu_J$ is introduced by the junctions. Next, we insert $\nabla\mu$ in Boltzmann transport equations and calculate $S$ for the system of RQW and junctions, expressing it as $S = S_0 + S_J$, where $S_0$ is Seebeck coefficient of RRWQ and $S_J$ is introduced by the junctions. Finally, the reduced figure of merit $ZT/Z_0T$ is calculated and $\mu$ dependences are presented for such traditional thermoelectric materials as Si and Ge. Analysis was made within the parabolic bands approximation and the abrupt junction's approximation. Since only heavy QSD doping was considered, we neglected the hole contribution in this transport.

## 2. Charge and heat transport in the RWQ with junctions

Cross-section of the system of RQW and periodic junction is shown in figure 2. We assume that there is a temperature gradient $\nabla T$ in the Y-dimension. Consequently, depletion depth depends on the Y-coordinate, and geometry factor gradient $\nabla G$ appears in the Y-direction. Presence of $\nabla G$ and $\nabla T$ modifies the electron distribution function and causes electron motion from the hot side to the cold side. This motion is compensated by thermoelectric voltage. Let us write Boltzmann transport equations [12] for the system of RQW and periodic series of junctions

$$J = \mathcal{L}^{11}\left(\mathcal{E} + \nabla\mu/e\right) - \mathcal{L}^{12}\nabla T \text{ and } J^Q = \mathcal{L}^{21}\left(\mathcal{E} + \nabla\mu/e\right) - \mathcal{L}^{22}\nabla T . \tag{5}$$

Here, $J$ is the electric current density, $J^Q$ is the heat current density, $\mathcal{L}^{ij}$ are coefficients, $\mathcal{E}$ is the electric field, and $e$ is the electron charge. Within the parabolic bands approximation, $\mathcal{L}^{ij}$ are the functions of integrals of type [13].

$$\Omega^{(\alpha)}(E) = \int_{-\infty}^{+\infty} dE\left(-\partial f_0/\partial E\right)\rho(E)\tau(E)\,\mathrm{v}_y^2\,(E-\mu)^\alpha \tag{6}$$

where $f_0$ is the electron distribution function, $\tau(E)$ is the electron lifetime, $\mathrm{v}_y$ is the electron velocity in the Y-direction, and $\alpha = 0, 1, 2$. Let us find how QSD affects $\Omega^{(\alpha)}(E)$. The QSD does not change dispersion relation and consequently $\mathrm{v}_y$. It reduces density of states (3D density) $G$ times, i.e. $\rho(E) = \rho_0(E)/G$ and increases the transport lifetime $G$ times [2], i.e., $\tau(E) = G\,\tau_0(E)$ (the latter fol-lows from Fermi's golden rule). Here, $\rho_0(E)$ and $\tau_0(E)$ are the density of states and carrier lifetime, respectively, in the case $a = 0$. Consequently, product $\rho(E)\tau(E)v_y^2$ in the RQW is the same as in the conventional quantum well of the same width (3D case) and in the bulk material. This product does not depend on $G$ and consequently it becomes independent of depletion depth. The QSD changes the distribution function $f_0$, since it increases $\mu$. The QSD influences integrals $\Omega^{(\alpha)}(E)$ by changing $\mu$ alone. Therefore, integrals are the same as in the bulk material having the corresponding chemical potential

$$\mathcal{L}^{ij} = \mathcal{L}_0^{ij}. \tag{7}$$

where $\mathcal{L}_0^{ij}$ are the coefficients of bulk material having the value of $\mu$ obtained by the conventional doping (for instance).

Further, we have to find $\mu$ for the system of RQW and junctions and insert it in (5) together with (7). Chemical potential of degenerated semiconductor ($-2 < \mu^* < 2$ where $\mu^* = \mu/k_B T$) can be written as [14] (note that frequently Fermi energy is used instead of chemical potential in semiconductor literature)

$$\mu = k_B T\left[\ln(n/N_C) + 2^{-3/2}(n/N_C)\right]. \tag{8}$$

where $n$ is the electron concentration and $N_C$ is the effective conduction band density of states. In the case of heavy QSD doping ($n \gg n_i$, where $n_i$ is intrinsic concentration), we can use (3) for $n$. Density of states reduces $G$ times in RQW, i.e. $N_C = N_{CB}/G$, where $N_{CB}$ is the CB effective density of states for bulk semiconductor. Inserting this and (3) in (8), we get

$$\mu = k_B T\left\langle \ln\left[n_{con}(G-1)/N_{CB}\right] + 2^{-3/2}\left[n_{con}(G-1)/N_{CB}\right]\right\rangle. \tag{9}$$

Figure 3 shows the reduced chemical potential dependence on geometry factor in RQW, plotted according to (9), for Si and Ge materials for different $n_{CON}$. The reason for choosing $n_{CON}$ values will be discussed in Section 4.

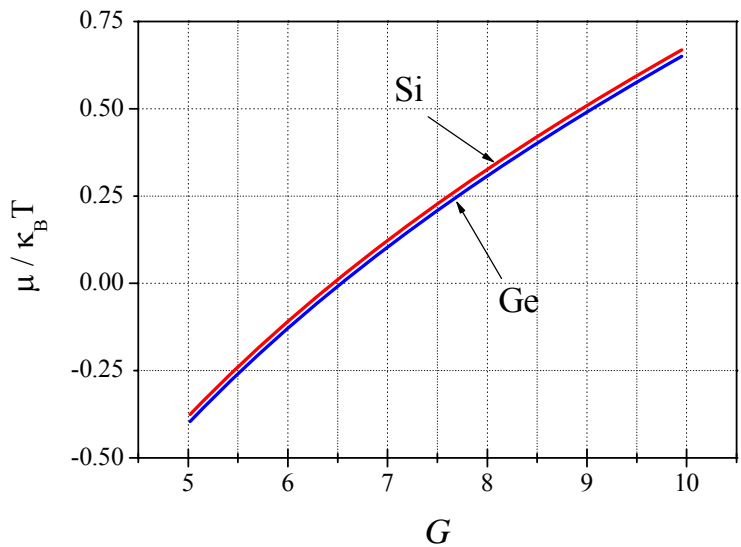

**Figure 3.** Chemical potential dependence on geometry factor in RQW for Si at $n_{CON} = 4.5\times10^{18}$ cm$^{-3}$ (red line) and Ge at $n_{CON} = 1.4\times10^{18}$ cm$^{-3}$ (blue line).

Further, we rewrite (9) as

$$\mu = \mu_0 + k_B T \left\langle \ln\left[(G-1)/(G_0-1)\right] + 2^{-3/2}\left(G-G_0\right) n_{con}/N_{CB} \right\rangle \qquad (10)$$

where $G_0$ is constant. Introduction of $\mu_0 \equiv \mu(G_0)$ defines reference material as n⁺-type semiconductor with electron concentration of $N_D = n_{CON}(G-1)$, or RRQW having constant geometry factor $(\partial G/\partial T) = 0$ and $G = G_0$. Next, we calculate the gradient of (10) taking into account that $N_{CB} \propto T^{3/2}$. The result is

$$\nabla\mu = \nabla\mu_0 + \theta\nabla T \ . \qquad (11)$$

where $\theta \equiv k_B \left[ \ln\dfrac{G-1}{G_0-1} - \dfrac{1}{2}\xi\left(G-G_0\right) + T\left(\dfrac{1}{G-1}+\xi\right)\dfrac{\partial G}{\partial T} \right] \qquad (12)$

and $\xi \equiv 2^{-3/2} n_{con}/N_{CB}$. Inserting (11) and (7) in Boltzmann equations, we find charge and heat currents in the system of RQW + J

$$J = \mathcal{L}_0^{11}\left(\mathcal{E} + \nabla\mu_0/e\right) - \left(\mathcal{L}_0^{12} - \mathcal{L}_0^{11}\,\theta/e\right)\nabla T \ \text{ and} \qquad (13a)$$

$$J^Q = \mathcal{L}_0^{21}\left(\mathcal{E} + \nabla\mu_0/e\right) - \left(\mathcal{L}_0^{22} - \mathcal{L}_0^{21}\,\theta/e\right)\nabla T \ . \qquad (13b)$$

Further, Seebeck coefficient, electrical conductivity, and electron gas heat conductivity can be found from (13a) and (13b) in a conventional way.

$$S = \left(\mathcal{L}_0^{12} - \mathcal{L}_0^{11}\,\theta/e\right)\Big/\mathcal{L}_0^{11} = S_0 - \left(\theta/e\right) \ , \qquad (14)$$

$$\kappa_e = \mathcal{L}_0^{22} - \mathcal{L}_0^{21}(\theta/e) - \left(\mathcal{L}_0^{12} - \mathcal{L}_0^{11}\,\theta/e\right)\mathcal{L}_0^{21}\Big/\mathcal{L}_0^{11} = \kappa_{e_0} , \qquad (15)$$

$$\sigma = \mathcal{L}_0^{11} = \sigma_0 . \qquad (16)$$

Electrical and thermal conductivity in the system of RQW and junctions remain unaffected (with respect to RRQW or bulk material having the same $\mu$ value) by a series of junctions, and $S$ change according to (14). To calculate $\theta$ and then $S$, we have to find $\partial G/\partial T$ first (12).

## 3. Geometry factor temperature dependence in RQW with p⁺–n⁺ and metal–n⁺ junctions

Depletion region reduces the effective height of the ridge from $a$ to

$$a_{eff}(T) = a - d(T) \ . \qquad (17)$$

Here, $d(T)$ is the depletion region depth. Differentiating (17), inserting (1), and taking into account that $G = G(T)$, we find

$$\left(\partial G/\partial T\right) = (dG/dT) = \left(G/a_{eff}\right)\left[d\,d(T)/dT\right]. \qquad (18)$$

Let us find $\partial G/\partial T$ in RQW with series of p$^+$–n$^+$ junctions first (figure 4). Within the abrupt junction approximation, $\mathrm{d}(T)$ has the following form [15]:

$$\mathrm{d}_{\mathrm{p-n}}(T) = \left[\frac{2\varepsilon_{\mathrm{S}}}{e}\frac{N_{\mathrm{A}}}{N_{\mathrm{D}}(N_{\mathrm{A}}+N_{\mathrm{D}})}\left(\varphi_{\mathrm{bi}}-\frac{2k_{\mathrm{B}}T}{e}\right)\right]^{1/2}. \qquad (19)$$

Here, $\varepsilon_{\mathrm{S}}$ is the dielectric permissibility of the material, $k_{\mathrm{B}}$ is the Boltzmann constant, $\varphi_{\mathrm{bi}} = \varphi_{\mathrm{bin}} + \varphi_{\mathrm{bip}}$ is the built-in potential, $N_{\mathrm{A}}$ is the acceptor concentration in p-type layer, and $N_{\mathrm{D}}$ is the QSD doping concentration.

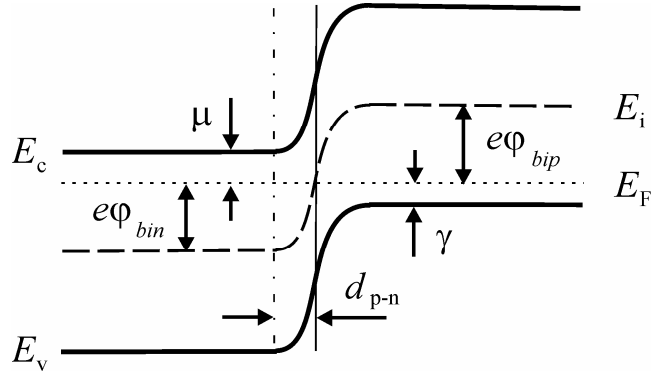

**Figure 4.** Energy diagram of p$^+$–n$^+$ junction at the top of the ridge. Ridge (left) is QSD-n doped and additional layer (right) is acceptor doped.

The built-in potential is (figure 4)

$$\varphi_{\mathrm{bi}} = \varphi_{\mathrm{bin}} + \varphi_{\mathrm{bip}} = \left(E_{\mathrm{g}} + \mu + \gamma\right)\big/e, \qquad (20)$$

where $E_{\mathrm{g}} = E_{\mathrm{C}} - E_{\mathrm{V}}$ and $\gamma$ is the chemical potential of p$^+$-type layer. For the case shown in figure 4, both $\mu$ and $\gamma$ are negative as $\mu$ is measured from the RQW conduction band bottom and $\gamma$ is measured from the p$^+$-type layer valence band top. Parameter $\gamma$ is determined using the formula similar to (8) for p$^+$-type semiconductor. Inserting (20) in (19) and introducing $\nu \equiv n_{\mathrm{con}}/N_{\mathrm{A}}$, we write

$$\mathrm{d}_{\mathrm{p-n}}(T) = \left\langle\frac{2\varepsilon_{\mathrm{S}}}{e^2 n_{\mathrm{con}}}\frac{(E_{\mathrm{g}}+\mu+\gamma-2k_{\mathrm{B}}T)}{(1-G^{-1})[1+\nu(1-G^{-1})]}\right\rangle^{1/2}. \qquad (21)$$

Differentiating (21) and inserting $d\mu/dT$ found from (9) and $d\gamma/dT$ found from formula analogical to (8), and further inserting result in (18) gives

$$(\partial G/\partial T)_{\mathrm{p-n}} = E_{\mathrm{p-n}}{}^{-1}\left[\left(\mu/T\right)+\left(\gamma/T\right)-5\,k_{\mathrm{B}}-(3/2)k_{\mathrm{B}}\,\xi(G-1)-(3/2)k_{\mathrm{B}}\,\delta\,\right]. \qquad (22)$$

where $\delta = 2^{-3/2}(N_{\mathrm{A}}/N_{\mathrm{V}})$, $N_{\mathrm{V}}$ is the valence band effective density of states in p$^+$-type layer, and $E_{\mathrm{p-n}}$ is the characteristic energy of the system of RQW and p$^+$–n$^+$ junctions and equals to

$$E_{\mathrm{p-n}} = (E_{\mathrm{g}}+\mu+\gamma-2k_{\mathrm{B}}T)\left[\frac{2\beta}{G}+\frac{G^{-1}}{(G-1)}+\frac{\nu\,G^{-1}}{G^{-1}+\nu(G-1)}\right]-k_{\mathrm{B}}\,T\left(\frac{G}{G-1}+\xi\right). \qquad (23)$$

where $\beta = a_{\text{eff}} / d$.

In the case of metal–n$^+$, Schottky junction will form at the top of the ridges (figure 5).

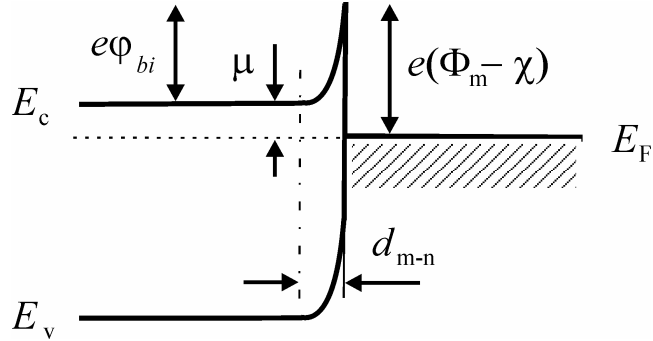

**Figure 5**. Energy diagram of metal–n$^+$ junction at the top of the ridge. Ridge (left) is QSD-n doped.

Within abrupt junction approximation, the metal–n$^+$ junction depletion layer depth is [15]

$$d_{\text{m-n}}(T) = \left[ \frac{2\varepsilon_S}{e} \frac{1}{N_D} \left( \varphi_{\text{bi}} - \frac{k_B T}{e} \right) \right]^{1/2}. \tag{24}$$

Here, $\varphi_{\text{bi}} = \Phi_m - \chi + \mu / e$ where $\Phi_m$ is the metal work function and $e\chi$ is the semiconductor electron affinity measured from the bottom of conduction band. Repeating the above-described steps for metal–n$^+$ junction, we found

$$(\partial G / \partial T)_{\text{m-n}} = E_{\text{m-n}}^{-1} \left[ (\mu / T) - (5/2) \, k_B - (3/2) k_B \, \xi (G-1) \right]. \tag{25}$$

Here, $E_{\text{m-n}} = (e\Phi_m - e\chi + \mu - k_B T) \left[ \dfrac{2\beta}{G} + \dfrac{G^{-1}}{(G-1)} \right] - k_B T \left( \dfrac{G}{G-1} + \xi \right). \tag{26}$

Investigation of (22) and (25) shows that both $(\partial G / \partial T)_{\text{p-n}}$ and $(\partial G / \partial T)_{\text{m-n}}$ strongly depend on $G$, and this can be used to increase Seebeck coefficient considerably. However, $(\partial G / \partial T)_{\text{p-n}}$ diverges and changes the sign for the value of $G$ for which $E_{\text{p-n}} = 0$ (the same happens in the case of metal–n$^+$ junction). Care should be taken to keep $G$ far enough from the divergence point and avoid the change of $(\partial G / \partial T)_{\text{p-n}}$ sign ($S$ will also change its sign). At the same time, $G$ should be chosen so that $-(\partial G / \partial T)_{\text{p-n}}$ is positive and high enough to obtain large enhancement in $S$ value. Additionally, $\gamma$ (acceptor doping of p-type layer) can be varied to attain the desired $(\partial G / \partial T)_{\text{p-n}}$ value. In the case of metal–n$^+$ junction, $\Phi_m$ can be varied instead of $\gamma$.

Equations (22) and (25) are obtained on the basis of (1) describing dependence of $G$ on the dimensions. However, (1) is valid only for a range of geometry factors $5 < G < 10$ and for $a \ll w$. Let us try to find $G$ for the arbitrary geometry. This requires solving of the time-independent Schrödinger equation in the ridged geometry [4]. Unfortunately, there is no analytical solution in the ridged well (solution contains infinite sums). However, there are fairly accurate numerical methods. Mathematically, there is no difference between QSD and the electromagnetic mode of depression, and Helmholtz equation and the same boundary conditions are used in both the cases. Helmholtz spectrum calculation can be found in the literature related to Casimir effect. Casimir energy exhibits strong dependence on the photon spectrum and consequently on the geometry of the vacuum gap [16]. A number of geometries, including double-side ridged geometry [17] and double-side corrugated geometry [18], were investigated. New, optical approach for arbitrary geometry was also developed [19]. Unfortunately, none of the above-described methods allows simple analytical solution like (1). Consequently, we have to accept the limit $5 < G < 10$ for further analysis. However, the above-listed numerical methods allow one to go well beyond those limits.

## 4. Seebeck coefficient of RQW with p⁺–n⁺ and metal–n⁺ junctions

It is reasonable to calculate Seebeck coefficient of RQW with junctions relative to RRQW, in which geometry factor is constant $(\partial G / \partial T) = 0$. This will allow comparison of dimensionless figure of merit $ZT$ with $Z_0 T$, using similarity of electric (16) and heat conductivities (15). In (10)–(12) $G_0$ is arbitrary. To simplify the comparison, we choose $G_0$ so that RWQ with junctions and reference RRQW have the same $\mu$ value. Equation of chemical potentials leads to $G_0 = G$ (9). Inserting this in (12) and further inserting the obtained result in (14), we obtain for Seebeck coefficient of RQW with junctions

$$S_{\text{p-n, m-n}} = S_0 - \frac{k_{\text{B}} T}{e} \left( \frac{1}{G-1} + \xi \right) \left( \frac{\partial G}{\partial T} \right)_{\text{p-n, m-n}} . \qquad (27)$$

Here, $S_{p-n}$ and $S_{m-n}$ are the Seebeck coefficients of RQW with p⁺–n⁺ and metal–n⁺ junctions, respectively. 3D Seebeck coefficient of reference RRQW do not differ from Seebeck coefficient of the bulk material, and within the parabolic bands approximation it is equal to [20]

$$S_0 = \frac{k_{\text{B}}}{e} \left[ \frac{r+5/2}{r+3/2} \frac{F_{r+3/2}(\mu^*)}{F_{r+1/2}(\mu^*)} - \mu^* \right] . \qquad (28)$$

Here, $r$ refers to scattering parameter and it is assumed that electron lifetime $\tau(E) \propto E^r$, $\mu^* = \mu / k_{\text{B}} T$ is the reduced chemical potential, and $F(\mu^*)$ are Fermi integrals. In the case of no impurities (QSD doping) and low energies, acoustic phonons are responsible for electron scattering and $r = 0$. It should be noted here that RQW has $G$ times more width [2] with respect to the conventional quantum well. As 3D $\rho(E)$ in wide quantum wells tends to $\rho(E)$ of the bulk material of the same width, using (28) for RRQW is a good approximation. However, in the case of thin layers oscillatory behaviour of transport coefficients should be considered [21].

To find the ratio $Z / Z_0$, we use relations (15) and (16) and obvious relation between lattice thermal conductivities $\kappa_l = \kappa_{l0}$, all together leading to

$$Z_{\text{p-n, m-n}} / Z_0 = \left( S_{\text{p-n, m-n}} / S_0 \right)^2 . \qquad (29)$$

Figure 6 shows the dependence $S_0(\mu)$ according to (28). The $S_{\text{p-n}}(\mu)$ in the same figure is determined by first, calculating $S_{\text{p-n}}(G)$ by inserting (22) in (27), and then (9) was used to convert $X$-axis so that $S_{\text{p-n}}(\mu)$ was obtained from $S_{\text{p-n}}(G)$. The ratio $Z_{\text{p-n}}(\mu) / Z_0(\mu)$ in the same figure is calculated according to (29). We present $\mu$ dependences as they allow the understanding of possible $\mu$ ranges, within which real devices can operate without changing sign of $(\partial G / \partial T)_{\text{p-n}}$ (22) and (23) and consequently $S_{\text{p-n}}$ sign.

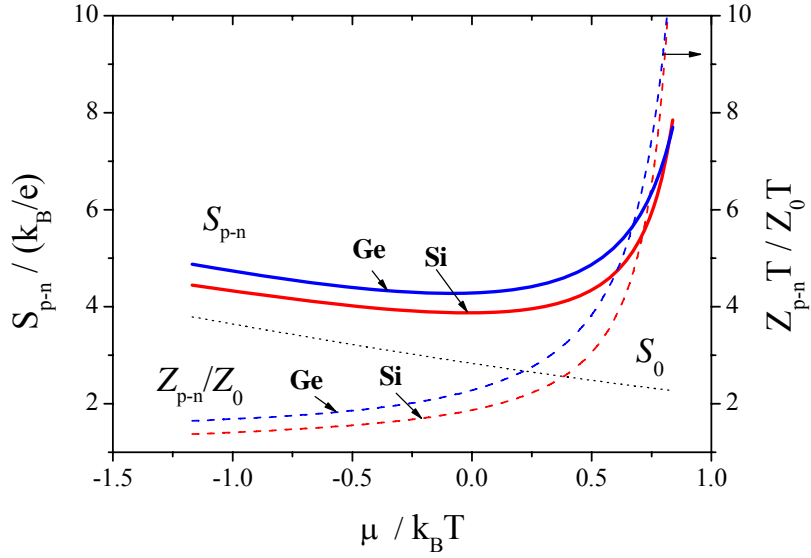

**Figure 6.** $S_{p-n}(\mu)$ – solid lines, $S_0(\mu)$ – dotted line, and $Z_{p-n}(\mu)/Z_0(\mu)$ – dashed lines (belong to right $Y$-axis). Dependences $S_{p-n}(\mu)$ and $Z_{p-n}(\mu)/Z_0(\mu)$ are calculated for Si and Ge materials.

Traditional thermoelectric materials Si and Ge are chosen as examples. Dependences are plotted for the following parameters: Material Si, $n_{CON} = 4.5 \times 10^{18}\,\mathrm{cm}^{-3}$, $a = 20$ nm, $\gamma = 3.3$, $\beta = 0.1$, $T = 300$ K; Material Ge, $n_{CON} = 1.4 \times 10^{18}\,\mathrm{cm}^{-3}$, $a = 35$ nm, $\gamma = -1$, $\beta = 0.25$, $T = 300$ K. For both Si and Ge, we choose $n_{CON}$ so that, for value $G=10$, $\mu$ was close to the optimum value $\mu_{OPT}/k_B T = r + 1/2$ [19]. Figures 5 and 3 allow finding of $S_{p-n}(G)$ and $Z_{p-n}(G)/Z_0(G)$ dependences as well. For this, $\mu$ should be determined from the desired point in figure 5 and next $G$ can be found using figure 3. Next, $L$ can be found by first determining d from $n_{CON}$ and $G$ values using (21) and next, determining $a_{eff}$ from (17) and inserting it in (1) instead of $a$. For the above parameter values and $G = 10$, we got $L = 18$ nm for Si and $L = 75$ nm for Ge. It should be noted here that RQW exhibits quantum properties at $G$ times more widths with respect to the conventional quantum well [2].

Figure 6 shows large enhancement of figure of merit in p$^+$–n$^+$ junction RQW both for Si and Ge materials. The dependences are quite identical for materials having rather different band gaps (1.12 eV for Si and 0.66 eV for Ge [22]). This shows that despite $E_g$ entering expressions for depletion depth (21) and $(\partial G/\partial T)_{p-n}$ (22), almost similar $S_{p-n}(\mu)$ and $Z_{p-n}(\mu)/Z_0(\mu)$ dependences can be obtained by matching such parameters as $\beta$ and $\gamma$.

Figure 7 shows dependences in the case of metal–n$^+$ junctions. The $S_{m-n}(\mu)$ was determined by first calculating $S_{m-n}(G)$ by inserting (25) in (27) and then using (9) for $X$-axis to convert $S_{m-n}(G)$ to $S_{m-n}(\mu)$. Both $S_{m-n}(\mu)$ and $Z_{m-n}(\mu)/Z_0(\mu)$ dependences are similar for Si and Ge. Two curves are not distinguishable on the given scale.

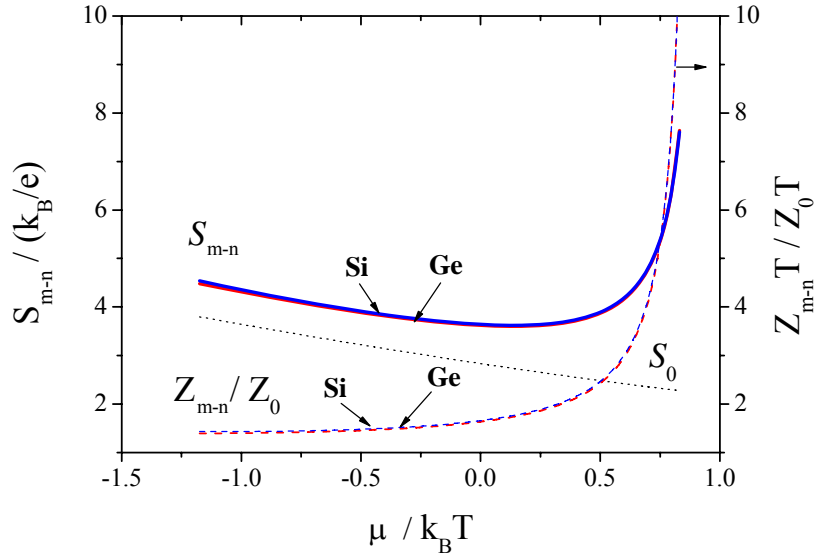

**Figure 7**. $S_{m-n}(\mu)$ − solid lines, $S_0(\mu)$ − dotted line, and $Z_{m-n}(\mu)/Z_0(\mu)$ − dashed lines.

Curves are plotted for the following parameters: Material Si, $n_{CON} = 4.5 \times 10^{18}\,\mathrm{cm}^{-3}$, $a = 15$ nm, $\beta = 0.55$, $(\Phi_m - \chi) = 11.3\ k_B T / e$, $T = 300$, K; Material Ge, $n_{CON} = 1.4 \times 10^{18}\,\mathrm{cm}^{-3}$, $a = 28$ nm, $\beta = 0.4$, $(\Phi_m - \chi) = 15\ k_B T / e$, $T = 300$ K. RQW width $L$ determined in the above-described way are $L = 60$ nm for Si and $L = 85$ nm for Ge.

As the analysis shows, high thermoelectric figure of merit can be obtained in both $p^+n^+$ junction and metal–$n^+$ junction cases. Metal–$n^+$ junction RQW seems to be simple in fabrication when compared with $p^+–n^+$ junction RQW, since metal film deposition is more straightforward than epitaxial growth of the semiconductor layer. However, $p^+–n^+$ junction allows more precize regulation of $(\partial G / \partial T)$, as it can be done by matching the acceptor concentration ($\gamma$ value), which is less complex than finding metal with required work function ($\Phi_m$ value in the case of metal-$n^+$ junction).

## 5. Conclusions

Thermoelectric transport coefficients were investigated in the system of ridged quantum wells and periodic series of $p^+–n^+$ and metal–$n^+$ junctions at the top of the ridges. Analysis was made on the basis of Boltzmann transport equations. It was shown that the Seebeck coefficient increases considerably. At the same time, electrical and thermal conductivities remain unaffected by the series of junctions. This allows large enhancement of thermoelectric figure of merit. Dependence of Seebeck coefficient on geometry factor $G$ and junction parameters was investigated and the analytical expression was obtained. Seebeck coefficient changes sign for some value of $G$. Dependences of $S$ and $ZT$ on chemical potential were presented for both $p^+–n^+$ and metal–$n^+$ junction RQW (separately for Si and Ge materials). Calculations show one order of magnitude increase in thermoelectric figure of merit with respect to the bulk material.

## Acknowledgment

The author thanks the Physics Department of New York University, where part of this work was done, for their hospitality.